\documentclass[manuscript]{acmart}
\AtBeginDocument{%
  }

\setcopyright{acmlicensed}
\copyrightyear{2025}
\acmYear{2025}
\acmDOI{XXXXXXX.XXXXXXX}

\usepackage{adjustbox}
\usepackage{caption}
\usepackage{subcaption}

\begin{document}

\title{Privacy-Driven Network Data for Smart Cities}

\author{Tânia Carvalho}
\email{tmcarvalho@tekprivacy.com}
\orcid{0000-0002-7700-1955}
\affiliation{%
  \institution{Faculdade de Ciências da Universidade do Porto and TekPrivacy}
  \city{Porto}
  \country{Portugal}
}

\author{José Barata}
\affiliation{%
  \institution{Associação Porto Digital}
  \city{Porto}
  \country{Portugal}}
\email{larst@affiliation.org}

\author{Henish Balu}
\affiliation{%
  \institution{Associação Porto Digital}
  \city{Porto}
  \country{Portugal}
}

\author{Filipa Moreira}
\affiliation{%
 \institution{TekPrivacy}
 \city{Porto}
 \country{Portugal}}

\author{João Bastos}
\affiliation{%
  \institution{Associação Porto Digital}
  \city{Porto}
  \country{Portugal}}

\author{Luís Antunes}
\orcid{0000-0002-9988-594X}
\affiliation{%
  \institution{Faculdade de Ciências da Universidade do Porto and TekPrivacy}
  \city{Porto}
  \country{Portugal}}

\renewcommand{\shortauthors}{Carvalho et al.}

\begin{abstract}
A smart city is essential for sustainable urban development. In addition to citizen engagement, a smart city enables connected infrastructure, data-driven decision making and smart mobility. For most of these features, network data plays a critical role, particularly from public Wi-Fi infrastructures, where cities can benefit from optimized services such as public transport management and the safety and efficiency of large events. 
One of the biggest concerns in developing a smart city is using secure and private data. This is particularly relevant in the case of Wi-Fi network data, where sensitive information can be collected. This paper specifically addresses the problem of sharing secure data to enhance the quality of the Wi-Fi network in a city. Despite the high importance of this type of data, related work focuses on improving the safety of mobility patterns, targeting only the protection of MAC addresses. On the opposite side, we provide a practical methodology for safeguarding all attributes in real Wi-Fi network data.
This study was developed in collaboration with a multidisciplinary team of legal experts, data custodians and technical privacy specialists, resulting in high-quality data.
On top of that, we show how to integrate the legal considerations for secure data sharing.
Our approach promotes data-driven innovation and privacy awareness in the context of smart city initiatives, which have been tested in a real scenario.

\end{abstract}





\maketitle

\section{Introduction}
%% SMART CITY

Many critical problems in cities, especially the current overwhelming urbanization, can be solved with advanced technologies and methods that can make cities smarter. Such problems include, for example, traffic congestion, environmental pollution, and crowds. By definition, a smart city leverages the information and communication technology infrastructure, human resources, social capital, and environmental assets to promote economic development, social/environmental sustainability, and improve the overall quality of life of its citizens~\cite{silva2018towards}.

% Wi-Fi NETWORK DATA
In smart cities, Wi-Fi network data is a critical component of the information and communication technology infrastructure, enabling real-time monitoring of urban dynamics. Wi-Fi network data is the digital information transmitted over a wireless network using Wi-Fi technology~\cite{lehr2003wireless}. Cities have an increasingly dense network of connected access points (APs) that enable people to access the internet on a range of different devices. 
When a smartphone, laptop or tablet (moving objects) connects to a Wi-Fi network, it establishes a wireless connection with a nearby AP. These mobile devices, characterized by their portability, often move through environments and may connect to multiple APs over time, enabling continuous wireless communication across spatially distributed network nodes.

A trace, produced by a moving object, is typically represented as a temporal sequence of spatial points, each associated with a corresponding timestamp. Although structurally simple, such traces encapsulate rich contextual information about human behavior and urban dynamics. The analysis and mining of these traces can provide precise insights and uncover latent patterns or knowledge related to both individuals and the city environment~\cite{pan2013trace, Roux2017ApproachesAT}.

% PRIVACY
Despite the significant benefits of mining movement traces, such data can contain semantic information that may reflect individual preferences, social interactions, and specific physical locations. Disclosure or misuse of this information poses potential risks ranging from perceived invasion of privacy and psychological discomfort to inadvertent exposure of personal activities, which in extreme cases could lead to tangible threats to an individual's safety and well-being~\cite{krumm2009survey}. Moreover, under privacy regulations, Media Access Control (MAC) addresses are classified as personal data according to Art. 4 of the General Data Protection Regulation (GDPR). Therefore, within this research context, privacy concerns are primarily associated with the collection and processing of location and movement data, emphasizing the importance of individuals' ability to control access to and dissemination of their location-specific information.

% PROBLEM OF DATA SHARING 
For higher efficiency and accessibility in cities, digital inter-connectivity plays an important role. Thus, multiple parties aim to share and access pertinent data where the joining of two or more data sets with a common subject of interest should not compromise individual privacy. %In addition to sharing data between different stakeholders, this type of data is also very relevant to open data science.
However, many privacy challenges have been raised in the development of smart cities concerning data sharing practices, especially on tracking trajectories~\cite{braun2018security}.
Coupled with this, a concern is that most people are unaware that their Wi-Fi is a potential source of tracking~\cite{demir2013wi}. In particular,~\citet{ten2019privacy} show that people are not aware of the purposes of Wi-Fi tracking, and a curious conclusion is that the mean scores of privacy are higher for women than for men, indicating that the group of females tend to have more worries regarding their privacy. This result highlights the high importance of raising consciousness on the harms of privacy invasions.

% OBJECTIVES
As such, in this paper, we aim to: \textit{i)} raise awareness on privacy concerning movement objects; \textit{ii)} provide privacy-preserving methods for secure sharing of real Wi-Fi network data, and \textit{iii)} show how the participation of each member of a multidisciplinary team improves privacy, maintains utility and enhances quality of the data.

These objectives led to the following contributions.

% CONTRIBUTIONS
\begin{itemize}
    \item We demonstrate an (pseudo) anonymization procedure covering various data attributes—beyond MAC addresses—in real Wi-Fi network data;
    \item We quantify privacy risk for certain subsets along with a data utility assessment;
    \item We allow the sharing of a secure version of the Wi-Fi network data with multiple stakeholders or as open data;
    \item We translate legal requirements into actionable technical guidelines for responsible data sharing.
\end{itemize}

This work was developed in the context of a hackathon, in which the objective was to explore the potential of open city data to develop solutions that may impact the community using Wi-Fi network data. 
Such initiatives highlight the critical need for explicit guidance on privacy-preserving approaches to enable secure data sharing, thereby accelerating the generation of insights without compromising privacy. Most importantly, it demonstrates the practical applicability of our methodology in the real world. To the best of our knowledge, we are the first in presenting all the steps of de-identification process in type of data along with legal considerations.

% STRUCTURE
The remainder of the paper is organized as follows. Section~\ref{sec:sota} provides a literature review on data privacy principles with a focus on the protection of Wi-Fi network data. The process of privacy preservation along with the evaluation of privacy risk and utility is presented in Section~\ref{sec:methodology}. Section~\ref{sec:discussion} provides a thorough discussion on such results and conclusions are presented in Section~\ref{sec:conclusion}.

\section{Literature Review}~\label{sec:sota}

In the following section, we provide concise notions of relevant background knowledge on data protection mechanisms in the context of data-sharing practices suitable for several tabular data domains. We also explore relevant work on privacy preservation of Wi-Fi network data. Finally, we summarize our contributions to the current state-of-the-art.

\subsection{Notions on Data Privacy}
We first present specific privacy challenges and risks associated with data in smart city environments, followed by general guidelines on common considerations for the sharing and release of data.

\paragraph{\textbf{Smart Cities Data.}}
% The majority of data in smart cities concerns movement data, often called trajectory data, which refers to records of objects or individuals moving through space over time -- spatiotemporal data.

A relevant application domain within smart cities is mobility analysis which often involves the use of trajectory data, a type of spatiotemporal data that captures the movement of entities (e.g. individuals, vehicles or objects) through space over time.

According to~\citet{finn2013seven}, the privacy of spatiotemporal data implies that individuals have the right to move through public or semi-public spaces without being identified, tracked, or monitored. This conception of privacy holds significant social value: when citizens are able to navigate public spaces freely and without the fear of surveillance, they are more likely to experience a sense of democratic freedom and personal autonomy.

In the context of Wi-Fi network data, connections to public Wi-Fi networks can raise notable privacy concerns. While mobile devices turn on Wi-Fi, they transmit Wi-Fi signals to be connected to a Wi-Fi access point (AP). Thus, it becomes possible to construct highly detailed profiles of their movements and behaviors over time, often without their explicit awareness or consent. Wi-Fi tracking can provide information on human dynamics such as the people's paths, the size of the crowd, the duration and frequency of visits~\cite{michael2013location}. This type of information can be used for several malicious purposes. For instance, unsolicited advertisements from shops when a mobile user approaches, firms using location information to impose strict performance measures on employees, and even dangerous or repressive, like criminals determining the right time to intrude on the individual’s house.

\paragraph{\textbf{Data Sharing Privacy.}}
Sharing data fosters collaboration between organizations, or even between departments within the same organization, but also enhances scientific progress when shared publicly. It enables reproducibility, reduces redundant effort and supports transparency and accountability.

The vast data-sharing practices and their benefits have led many organizations, institutions and governments to endorse open data initiatives. For example, the European Union~\cite{eudirective}, and the US Chief of Information Officer's Council~\cite{act2018}, among others, encourage governments to provide access to public sector information, promoting its reuse and accelerating innovation.~\citet{attard2015systematic} presents the open government data life cycle along with guidelines for publishing data. Concerning open research data, it often adheres to the FAIR (Findable, Accessible, Interoperable and Reusable) guiding principles~\cite{wilkinson2016fair, fair_sharing}. FAIR describes exploration, sharing, and reuse considerations in data publishing.

Concerning data collection, this process should follow the Privacy by Design and Privacy by Default approach~\cite{privacy_by_design}. According to Art. 25 of the GDPR, privacy by design means that personal information must be protected in any given IT system. On the other hand, privacy by default means that only the necessary personal data is processed to achieve specific purposes. This approach enables organizations to optimize data storage efficiency, improve sustainability, and enhance their environmental footprint~\cite{alessi2021privacy}.

% DPIAs, FIVE safes and SDC
Data Protection Impact Assessments (DPIAs) are an integral component of data privacy management. Under Art. 35 of GDPR, a DPIA is required when a project involves a potentially high risk to the rights and freedoms of data subjects. DPIA is a procedure that should be used as an early cautioning framework, as it identifies potential privacy infringements associated with data processing activities. This framework considers the full data life cycle, from collection to sharing, and assesses how different elements, such as technology and organizational practices, contribute to the overall risk.
Thus, designing a safe protocol is essential when publishing data sets, and the Five Safes framework~\cite{arbuckle2019five, arbuckle2020building} provides a suitable approach. This framework is usually used as an auxiliary to decision-making that helps assess the privacy risks and potential benefits of releasing or sharing data. These protocols are used to analyze privacy risks at different stages of the information flow. 

To ensure \textit{safe data} for secure data sharing, privacy mechanisms must be applied to protect individuals' personal information and provide sufficient granularity for useful and meaningful further data analysis. One such mechanism is the de-identification process. In statistical disclosure control, data transformation using various Privacy-Preserving Techniques (PPTs) is key to ensuring the protection of the data when released (i.e., de-identified)~\cite{carvalho_survey}.
A \textit{safe output} is a set of statistics, such as descriptive statistics, that are unlikely to reveal any personal information about a data subject. Requirements and guidelines have been discussed for checking statistical outputs to ensure they are safe to be released~\cite{outputs_check, safe_outputs} in which the de-identification process has been used to facilitate acceptable statistical outputs. 

% de-identification process
The de-identification process consists of three main phases~\cite{carvalho_survey}:
\textit{i)} raw disclosure risk and data utility assessment, \textit{ii)} application of privacy-preserving techniques mainly guided by the disclosure risk and attribute characteristics, 
and \textit{iii)} re-assessment of disclosure risk and data utility. 
If the balance between these two measures is not met, further refinement of the PPTs may be necessary. This iterative process is essential for successful de-identification.
% QIs
Moreover, its effectiveness heavily depends on the assumptions made about an attacker's background knowledge, particularly relevant in the selection of quasi-identifiers (QIs), such as date of birth, gender, and occupation. When combined, these QIs can form a unique signature, heightening the risk of personal information disclosure. Therefore, applying appropriate transformations to QIs to mitigate such risks is essential. Regarding the direct identifiers, such as names or social security number, are removed or replaced by a pseudonym.

Traditional techniques for transformation include generalization (recoding values into broader categories), suppression (replacing values with \textit{NaN} or special character) and noise to ensure a desired level of privacy. These can be validated using various tools, such as $k$-anonymity~\cite{samarati2001protecting} or differential privacy~\cite{dwork2008differential}. 
A data set is $k$-anonymous if each individual in a data set is indistinguishable from at least $k - 1$ other individuals. 
%On the other hand, 
To achieve differential privacy, the statistical results of a dataset should not be affected by the contribution of any individual. These strategies specifically address the problem of homogeneity and linkage attacks, several other methods exist regarding different types of attacks~\cite{jin2022survey,sampaio2023collecting}.
% k-anonymity, record linkage
The transformed data set is then evaluated in terms of its privacy risk, of which re-identification poses the most significant threat~\cite{wp29}. Two standard measures for re-identification risk are $k$-anonymity and record linkage. $K$-anonymity~\cite{samarati2001protecting} indicates how many $k$ occurrences occur in the data set for a given combination of QI values. An attacker can single out an individual if $k = 1$. On the other hand, record linkage (or linkability)~\cite{fellegi69} aims to measure the ability of re-identification by linking two records using similarity functions. 

\subsection{Network Data Privacy}
Tracking movement patterns of a person based on Wi-Fi connections is not a new topic, and several applications have been analyzed. For instance,~\citet{bonne2013wifipi} used the real-time gathered data for crowd control in a music festival. The authors were able to monitor the density of the crowd, analyze the flow patterns, how long people stay at the festival, and assess the audience sizes for different performances. In addition to cultural events, the educational domain has also been investigated. In particular, the access to \textit{Eduroam}~\footnote{https://eduroam.org}, an international Wi-Fi for students and researchers, has been subject to analysis of movement patterns.
~\citet{zhu2015measurement} found that students generally have a regular diet only on weekdays and have tardiness behavior on weekday mornings just by studying campus and dormitory Wi-Fi user distribution.~\citet{danalet2016location} used Wi-Fi traces to measure catering choices on a certain campus, then forecasting the average number of visits after the opening of a new self-service.~\citet{kalogianni2015passive} used passive Wi-Fi monitoring occupation and movement in campus buildings to improve future use of the campus. Concerning commercial purposes, Wi-Fi traffic can reveal information on shopping patterns, customer loyalty, dwell times, walking paths, real-time heatmaps, and even customer gender and age. Thus, Wi-Fi deployments can become powerful tools for conducting market research and gauging customer insights~\cite{redondi2018building}.

Although these studies are relevant to improving services in a smart city, the uniqueness of location traces can easily lead to the identification of individuals.~\citet{sapiezynski2015tracking} show that any application can use Wi-Fi permission to link users to other public and private identities, using data from social networks like Twitter or Facebook and geo-tagged payment transactions. Such cross-linking allows one to conclude that Wi-Fi scans should be considered a highly sensitive type of data. This problem was amplified during the global COVID-19 pandemic~\cite{smidt2021challenge,azad2020first,zakaria2022analyzing,zaidi2022differentially}.

Given such concerns, Wi-Fi based data should be transformed through appropriate PPTs to be further analyzed and linked with other data. In this regard,~\citet{demir2013wi} built a scenario in which an attacker could retrieve the original MAC address. For that, the authors anonymize the MAC address using different cryptographic hashing functions~\cite{menezes2018handbook}. Their results show that there is a trade-off between the protection of privacy and the feasibility of a long-period tracking system. Several other researchers have proposed privacy-preserving strategies for Wi-Fi signals~\cite{rusca2024privacy,demir2014analysing,garroppo2025enhancing,pang2009wifi}.

Although MAC addresses are masked, other concerns arise, such as the vulnerability of smaller groups within a particular area or event, which increases the attention of attackers, making it easier to correlate behavior patterns, device characteristics or movement trajectories. This problem is magnified in sparsely populated contexts or when unique device behaviors can be observed.
In a visiting frequency estimation,~\citet{ackermann2023privacy} purpose to remove data points when there is little activity. Such periods include nights when the streets are less busy and periods when fewer than 10 unique MAC addresses. Although this approach is promising as it allows for the protection of small groups, the authors discuss attack models that can be used to identify and track devices based on their Wi-Fi probe requests; we are interested in secure data sharing to improve the service of Wi-Fi in the city, and as such we aim to protect other data attributes as well.

The statistics per AP are crucial to the improvement of the Wi-Fi network in a smart city. Additionally, such statistics, while of limited use due to reporting granularity, are also
useful for crowd mapping, allowing learning building-level and floor-level occupancy counts. In this regard,~\citet{zaidi2022differentially} uses differential privacy to protect simple queries such as the number of individuals present in a particular building. %Nevertheless, introducing noise via differential privacy typically destroys data utility~\cite{carvalho2023towards}.

After de-identification,~\citet{Roux2017ApproachesAT} show that it is possible to perform several data analyses in anonymous Wi-Fi location-based data. The authors identify groups that reflect the order and composition of crowds at the football game -- the FA Cup Final.  
Although the data contains MAC addresses, one timestamp and one set of geographical coordinates (latitude and longitude), only the MAC is transformed, in particular anonymized. Since other attributes remain intact, this can justify the high precision of the results. However, the authors focus on large crowds while we aim to study the possibility of isolating small groups. % Also, we show the process of de-identification applied by transforming several attributes.

\subsection{Summary}
%Wi-Fi-based tracking systems have recently gained attention, especially during COVID-19 outbreak~\cite{smidt2021challenge,zakaria2022analyzing}. 
By collecting radio signals emitted by Wi-Fi devices, those systems can be effectively used to monitor crowd density and enhance safety management in public spaces such as urban parks and large-scale events.
Numerous studies have demonstrated the feasibility of tracking movement patterns using Wi-Fi signal data~\cite{bonne2013wifipi,danalet2016location,redondi2018building}.
These pattern movements are often drawn using the MAC addresses, which allows tracking individual devices and thus, their users. If retailers and businesses, for instance, have high expectations for physical tracking, it may pose a threat to citizens' privacy~\cite{demir2013wi,sapiezynski2015tracking,smidt2021challenge}.
%Hence, it is required careful consideration of privacy implications and mitigation strategies for MAC addresses.
Given the important role of Wi-Fi signals in enabling movement pattern analysis and enhancing various services, along with the need to safeguard user privacy, several studies have shown privacy protection mechanisms for this type of data~\cite {demir2013wi,rusca2024privacy,demir2014analysing,garroppo2025enhancing,pang2009wifi}. However, the majority of these works focus solely on protecting the MAC address.

Since we are interested in sharing data with more information besides the MAC address, it is crucial to provide mitigation strategies for other data attributes.
~\citet{braun2018security} have particularly highlighted the challenges for privacy preservation with high-dimensional data and show trustworthy data sharing practices. However, the authors show approaches for general secure data sharing, while we focus on the Wi-Fi network data domain.
Like~\citet{Roux2017ApproachesAT}, our interest lies in maximizing the potential of Wi-Fi network data to support a range of different applications such as urban planning, crowd management, public security, transportation optimization and advertisement. Although the authors provide an analysis of data after de-identification, only MAC address is transformed; other attributes must be equally analyzed in terms of privacy.
In our approach, we also analyze the privacy implications of small groups, and to protect them, we remove critical data points~\cite{ackermann2023privacy}.

Synthetic-based solutions have recently gained prominence as an alternative to traditional approaches. Examples include Generative Adversarial Networks (GANs) and interpolation methods~\cite{figueira2022survey}. For high privacy guarantees, differential privacy has been integrated into the synthetic data generation process (e.g. DPGAN~\cite{xie2018differentially}).
Despite their vast applications and how these approaches can circumvent many privacy aspects, such as, specifying the critical attributes, they may not be so suitable for this type of task. Synthetic data generation models often require too much time to train and is computational costly~\cite{carvalho2022differentially}. In our context, these methods may require substantial training time due to the large volume and high dimensionality of Wi-Fi network data. Thus, we do not account for this type of protection methods.

Besides the goal of protecting the whole dataset for secure data sharing, we want to create a set of metrics, for instance ``number of sessions for a given AP''. As such, we resort to traditional PPTs since we aim to maintain the truthfulness of data after transformations which is suitable for both objectives.

In summary, unlike related work that focuses on MAC address protection, we transform other Wi-Fi-related attributes to enhance privacy. We also perform a data privacy and utility analysis to demonstrate the effectiveness of the selected PPTs. Most importantly, beyond the technical implementation and details, we incorporate legal considerations to support the scientific community in conducting a DPIA. Our goal is to translate the current regulatory frameworks into simple steps that facilitate the development of privacy-conscious smart city innovations.

\section{Privacy-Preserving Wi-Fi Network Data}~\label{sec:methodology}
%This section provides the methodology details for safeguarding the dataset. We first describe the data and then, we demonstrate the efficiency of our methods in both privacy and utility.
Our methodology was first developed for data sharing in the Hack a City~\footnote{https://hackacity.eu} event in Porto, which gathered participants from all over the country, including public and private sectors. The main objective of the event was to raise awareness of privacy preservation among the participants, with the task of evaluating and improving the city's data.
Although data was shared with the participants under a confidential agreement, the de-identification process applied to this event data is also appropriate for public release or sharing among stakeholders, as it minimizes privacy risks to the greatest extent possible while preserving the data's utility.

To proceed with the de-identification process, it is essential to acknowledge that the team representing the data controller (data custodians and domain experts) is responsible for addressing the Five Safes framework~\cite{arbuckle2019five, arbuckle2020building} and conducting a comprehensive risk-benefit assessment. This preliminary analysis is important to assess the relevance of the study and the associated risks from data collection to publication, before starting any data transformation procedures. In summary, the benefits of data sharing for this particular study include: 
\begin{enumerate}
    \item simplification of the process of data availability, allowing rapid, secure access by other stakeholders, and
    \item apply the proposed solution to the new incoming data. 
\end{enumerate}
On the other hand, given the sensitive nature of network data, it is imperative to mitigate data protection risks while ensuring the fundamental right to privacy of users. Therefore, the establishment of a (pseudo) anonymization methodology becomes indispensable, allowing the risk of users' re-identification to be objectively quantified. Accordingly, the decision to release the data to the general public or to specific recipients is of great importance.

For an effective de-identification, we must understand the data domain and properties to apply the best Privacy-Preserving Techniques (PPTs) with an accurate parameterisation. As such, we subsequently describe the data, and finally, we present our approach for de-identification with an analysis on the effectiveness of the transformed data concerning both privacy and utility.

\subsection{Data}~\label{sec:data_description}
Typically, when we connect to a public network, like \textit{Eduroam}, the login credentials are sent to a RADIUS server. The RADIUS infrastructure ensures that the credentials are securely validated, and if authentication is successful, access is granted~\cite{florio2005eduroam}. This server generates logs for further network monitoring. Therefore, the provided data concern the logs for only three months, namely from July to October 2024, in the city of Porto in Portugal. The dataset comprises 28 attributes and 14.184.887 records.
As previous stated, we have two objectives: \textit{i)} securely share data and \textit{ii
)} create general metrics to assess and improve the quality and utility of the public Wi-Fi in the city. 
Concerning the latter, the following list of metrics is essential for the analysis.

\begin{itemize}
    \item Number of sessions (set of several connections, i.e, records within the same session identifier) for a given AP/hotspot (group of APs in the same general location)/set of APs/SSIDs in a flexible time window;
    \item Number of connections for a given AP/hotspot/set of APs/SSIDs in a flexible time window;
    \item Number of unique devices for a given AP/hotspot/set of APs/SSIDs in a flexible time window;
    %\item Number of new devices for a given AP/hotspot/set of APs/SSIDs in a flexible time window;
    \item Upload/Download for a given AP/hotspot/set of APs/SSIDs in a flexible time window;
    \item Session time for a given AP/hotspot/set of APs/SSIDs in a flexible time window.
    % \item Daily origin-destination matrices;
    %\item Number of unique devices, connections/sessions per institution/country of the \textit{Eduroam} network.
\end{itemize}

However, much information was collected which is not necessary for our goal; only a small part of the attributes is essential. As such, we only use 7 attributes for this purpose, which are described in Table~\ref{tab:data}. This dataset contains 6.800.830 sessions,  599.199 user devices and 581 APs.

\begin{table}[ht!]
  \caption{Description of Wi-Fi network data.}
  \label{tab:data}
  \begin{adjustbox}{max width=\textwidth}
  \begin{tabular}{ccl}
    \toprule
    \textbf{Attribute}&\textbf{Value}&\textbf{Description}\\
    \midrule
    \textit{acctsessionid} & string ID & Unique identifier for sessions in progress.\\
    \textit{acctstarttime} & date YYYY-MM-DD HH:MM:SS & Connection start timestamp.\\
    \textit{acctinputoctets} & numeric & Number of bytes sent by the user to the network.\\
    \textit{acctoutputoctets} & numeric & Number of bytes sent from the network to the user.\\
    \textit{acctsessiontime} & numeric & Number of seconds a session was active.\\
    \textit{callingstationid} & AA-BB-CC-DD-EE-FF & Login user's MAC address.\\
    \textit{calledstationid} & 11-22-33-44-55-66:Network Name (SSID) & MAC address of the AP receiving the connection request.\\
  \bottomrule
\end{tabular}
\end{adjustbox}
\end{table}

\subsection{Privacy-Utility Analysis}
Quantifying the risk of disclosure is a challenge since disclosure of confidential information generally occurs when an attacker has external information that the data controller often cannot anticipate. If an
attacker has more background knowledge than assumed, the risk of disclosure may be underestimated. Therefore, the controller needs to make prudent assumptions about such knowledge to predict the risk of disclosure. Typically, the controller determines the privacy risk under different scenarios (threat models), e.g. different sets of attributes that attackers may know. 

Attributes such as \textit{acctoutputoctets} and \textit{acctinputoctets}, which represent the volume of data sent and received by a device, may seem innocuous from a privacy perspective, but can actually pose privacy risks when included in Wi-Fi network datasets. For instance, when combined with other attributes such as timestamps, MAC addresses (even if pseudonymized), can reveal unique traffic volume patterns that may allow for re-identification of users or devices. Also, high or low data volumes during specific periods can suggest certain activities (e.g., video streaming, file transfers), potentially exposing sensitive user interactions. Given this, we use all attributes in our privacy analysis, assuming full background knowledge. We focus on critical groups, i.e., small groups of people given certain criteria.

\paragraph{\textbf{Initial Disclosure Risk Evaluation.}}
During the three-month period, 17\% of users connected to the network only once. This scenario is very risky. For example, \textit{user A} connects to \textit{AP1} at 12:05 PM. There’s only one record for \textit{user A} and it's tied to \textit{AP1} at a specific time. If there are only a few users who connect to \textit{AP1} at around 12:05 PM, \textit{user A} becomes easy to re-identify. There's a higher likelihood that this connection record is unique or can be cross-referenced with other data such as location data or time of day, to identify \textit{user A}.
%if \textit{user A} connects to \textit{AP1} at a given time, and only a small number of users are connected to the same AP at the same time, the limited set increases the risk of re-identification. Attackers can exploit temporal and spatial correlation to isolate and identify \textit{user A} when connecting to \textit{AP1}. As a mitigation measure, we remove cases where users connect only once.

There are also 0.7\% of critical APs, meaning they have few connections. If few users connect to a given AP, the identity of the user can be inferred by cross-referencing other sources using the timestamp and location. Despite the low probability, we remove such records. The minimum number of users sharing the same AP is now 114.

In addition, 5\% of users only connect to a single AP. A user who only connects to one AP can be easily traced because their data is unique and associated with a single location. For instance, a user connects to a coffee shop's Wi-Fi. If the shop has few customers or is located in a niche area, it becomes trivial to identify the user. For greater protection, we also remove these critical records. 

These transformations resulted in the removal of 1.4\% of the records, which corresponds to 201.769 observations. The dataset now contains 13.983.118 records. 
Despite the removal of the most critical scenarios, there is still another concern: the granularity of the original data. Attributes such as \textit{acctstarttime} and \textit{acctsessiontime} are presented in seconds and \textit{acctoutputoctets}/\textit{acctinputoctets} in bytes. The analysis of these attributes reveals a high level of risk. For example, more than 90\% of users have unique data usage patterns (upload/download). We therefore proceed with a transformation to reduce the level of information in these attributes.

\paragraph{\textbf{Privacy Preservation.}}
Given the sensitive nature of the dataset, we must transform all the attributes. Table~\ref{tab:transformations} summarizes the main PPTs applied to each attribute.

\begin{table}[ht!]
  \caption{Type of transformation applied to each attribute.}
  \label{tab:transformations}
  \begin{tabular}{cl}
    \toprule
    \textbf{Attribute}&\textbf{Privacy-Preserving Techniques}\\
    \midrule
    \textit{acctsessionid} & Pseudonyms generation. \\
    \textit{acctstarttime} & Suppression: removal of seconds and minutes. \\
    \textit{acctinputoctets} & Generalization: bytes to megabytes. \\
    \textit{acctoutputoctets} & Generalization: bytes to megabytes.\\
    \textit{acctsessiontime} & Generalization: seconds to minutes.\\
    \textit{callingstationid} & Pseudonyms generation.\\
    \textit{calledstationid} & Pseudonyms generation. \\
  \bottomrule
\end{tabular}
\end{table}

Given the high level of granularity in \textit{acctstarttime}, we start by transforming it into YYYY-MM-DD HH format, i.e. keeping only the information to the hour level instead of seconds. In the case of \textit{acctsessiontime}, which is numeric format, we just apply the operation of dividing each value by 60 to have the representation in minutes. To generalise \textit{acctinputoctets}/\textit{acctoutputoctets} to megabytes the apperation is $value / (1024^2)$. Regarding pseudonymisation, the values of the three attributes are replaced with pseudonyms that maintains the time structure. 

Since the dataset contains several SSIDs in \textit{calledstationid} where the majority corresponds to \textit{Porto. Free Wi-Fi}, a free public network maintained by Porto Digital, and \textit{Eduroam}, we further protect the remaining cases. In particular, we suppress SSIDs with fewer observations, keeping only the MAC address of the AP receiving the connection request in those observations.

\paragraph{\textbf{Effectiveness of Privacy-Preserving Techniques.}}
To evaluate the effectiveness of the transformations, excluding the generation of pseudonyms, we focus on the reduction of detail by using the $k$-anonymity~\cite{samarati2001protecting}. We compare the single outs, i.e. individuals that do not share the characteristics of the connection with anyone, before and after applying PPTs. In practice, this operation translates into grouping data by certain attributes and counting occurrences of sessions (\textit{acctsessionid}).
For instance, after suppressing the seconds and minutes of the \textit{acctstarttime}, the single outs for this attribute drop from 32.41\% to 0\%. A huge impact is noticed with the transformation of \textit{acctsessiontime}, which drastically reduced 76.07\%. Considering \textit{acctinputoctets} and \textit{acctoutputoctets}, the reduction was 25.75\% and 41.43\%, respectively. 

When combining these critical attributes, an important step in the $k$-anonymity analysis, the percentage of single outs increases, which is expected since we are assuming that an attacker has more knowledge. However, in network data or geolocated data, this approach is often not used. %This is because, as data contains many entries for the same \textit{acctsessionid} in different timestamps.
This type of data is dynamic and time-varying with unique patterns, which undermines the concept of forming stable equivalence classes, a characteristic of $k$-anonymity. Even after applying PPTs, the data may still be too specific to achieve anonymity, making it hard to form groups of size $k$. %Nevertheless, we stress the importance of this approach in finding critical scenarios.

% \begin{figure}[ht!]
%   \centering
%   \includegraphics[width=0.9\linewidth]{acmart-primary/transf_risk_final_risk_eduroam.png}
%   \caption{1907 Franklin Model D roadster. Photograph by Harris \&
%     Ewing, Inc. [Public domain], via Wikimedia
%     Commons. (\url{https://goo.gl/VLCRBB}).}
%   \Description{A woman and a girl in white dresses sit in an open car.}
% \end{figure}

In addition to this analysis, we note that per session, 20\% of users exceed the 75th percentile of data usage, which may still restrict the type of user activity in a given session. Also, 12.5\% of users accessed the network several times in a given hour. These users may be more vulnerable as their behavior may be linked to specific events. Furthermore, attackers can infer (with some ease) the end time of the session, which, if they have correctly mapped some APs, will allow them to calculate the trajectory of some users more efficiently. Another concern is that converting the \textit{acctsessiontime} from seconds to minutes may not be enough, as an attacker can easily reverse this transformation.
To avoid such risky scenarios, where an attacker can infer new information or reverse simple transformations, we add noise to the \textit{acctstarttime} attribute, setting it to a range of [-3,3] hours. This transformation is applied over the previous ones.

%\paragraph{\textbf{Utility Analysis}}
In terms of utility, we compare several of the metrics described in Section~\ref{sec:data_description}, before and after applying the aforementioned transformations. For simplicity, we present comparisons for a single attribute for each transformation that can impact utility, generalizing for similar metrics to avoid repeated calculations.

\begin{figure}[ht!]
    \includegraphics[width=15cm]{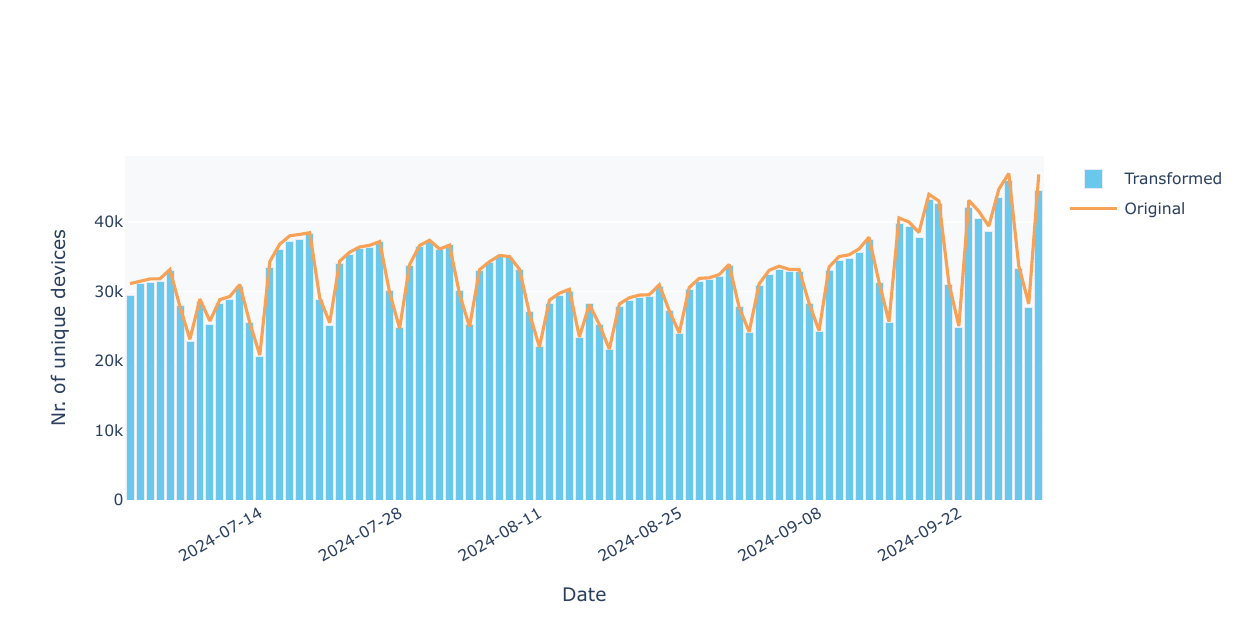}
    \caption{Number of unique devices per day before and after anonymization}
    \label{fig:devices_per_day}
    \Description{Comparison of device counts per day, before anonymization and after anonymization.}
    \centering
\end{figure}

To evaluate the impact of the suppression %and noise
applied to the \textit{acctstarttime} attribute, we examine the metric \textit{number of unique devices}, by first comparing the number of unique devices per day before and after the transformations, for the three-month period. This metric overlaps considerably with both \textit{number of connections} and \textit{number of sessions}, since the calculation method for the three attributes is very similar. 
Figure~\ref{fig:devices_per_day} shows the impact of such transformations concerning this metric.
As demonstrated, the suppression and noise have almost no impact on this metric when considering a single dimension (date). This can be attributed to the fact that the number of unique devices per day is almost exactly the same as the original, with a mean difference of approximately 325 devices; therefore, we can conclude that the utility trade-off is extremely low. 

\begin{figure}[ht!]
    \includegraphics[width=15cm]{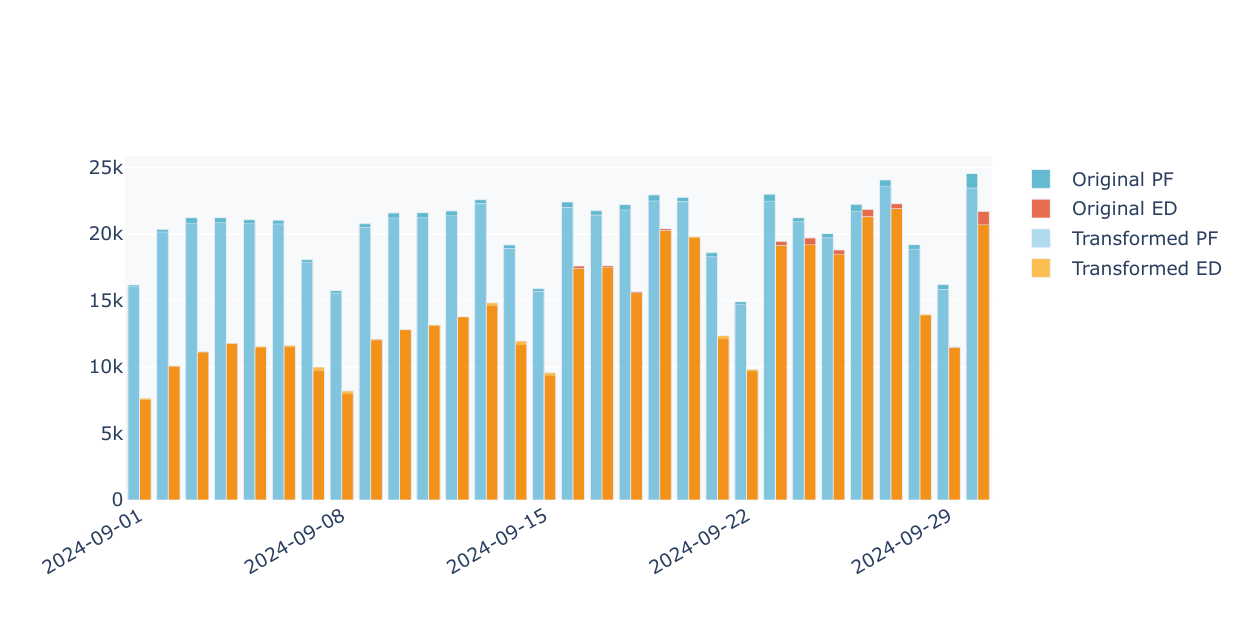}
    \caption{Number of unique devices per day given the \textit{Porto. Free Wi-Fi} (PF) \textit{SSID} and \textit{Eduroam} (ED) \textit{SSID} before and after anonymization.}
    \label{fig:devices_per_day_ssid}
    \Description{Comparison of device counts per day and SSID, before anonymization and after anonymization.}
    \centering
\end{figure}

Consequently, it is also very important to understand the effects of introducing another dimension of comparison, such as \textit{SSID}. Thus, Figure~\ref{fig:devices_per_day_ssid}
compares the number of unique devices per day and \textit{SSID} before and after the same transformations, specifically the \textit{Porto. Free Wi-Fi} \textit{SSID} and \textit{Eduroam} \textit{SSID}. The results indicate a good balance between privacy and utility, with no impact on the number of unique devices on either the \textit{Porto. Free Wi-Fi} \textit{SSID} or \textit{Eduroam} \textit{SSID}. In the majority of cases, the transformed values overlap with the original.

\begin{figure}[ht!]
    \includegraphics[width=15cm]{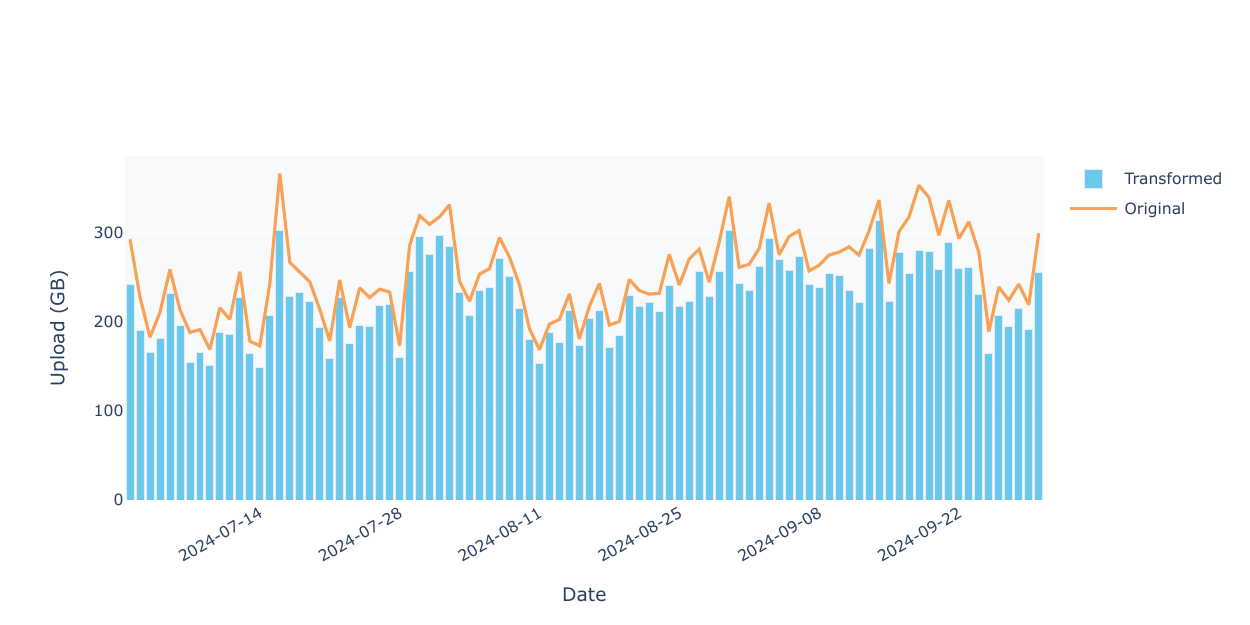}
    \caption{Total upload per day in gigabytes (GB) before and after anonymization.}
    \label{fig:upload_per_day}
    \Description{Total upload per day in gigabytes before and after anonymization}
    \centering
\end{figure}

Regarding the impact on the utility of the generalizations applied to the attributes \textit{acctinputoctets} and \textit{acctoutputoctets}, Figure~\ref{fig:upload_per_day} shows the total upload (\textit{acctinputoctets}) per day before and after the transformations.
% To evaluate the impact on utility of the generalizations performed on the attributes \textit{acctinputoctets}, \textit{acctoutputoctets} and \textit{acctsessiontime}, as well as the previously mentioned noise in \textit{acctstarttime} we compare the total upload per day before and after the transformations. 
% Once again, this comparison may be generalized to the other aforementioned second and third attributes, as the methods for calculation are identical. 
The results show that the transformed data follows the same pattern as the original. The total daily upload values are similar, although generally lower, with a maximum mean difference of around 27.7 gigabytes. In general, we observe that using truthfulness transformations, the ones that do not alter the data much, such as suppressing the seconds and minutes in the date, and generalize from bytes to megabytes, demonstrate a higher capacity for preserving the data utility.

\begin{figure}[ht!]
    \includegraphics[width=15cm]{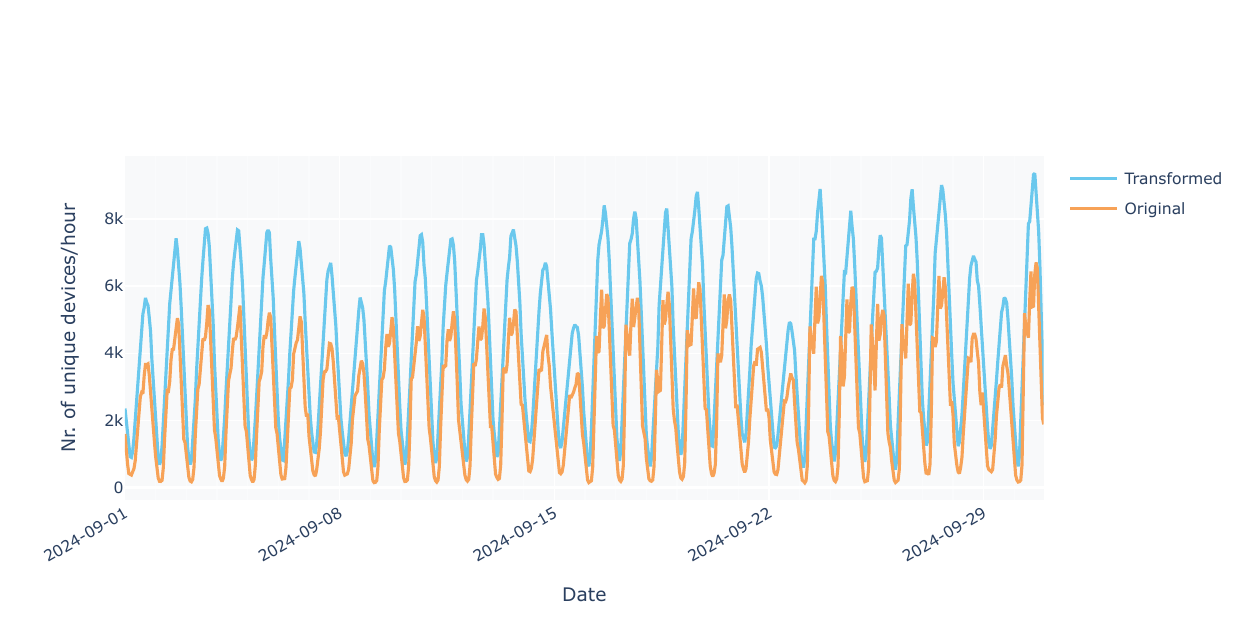}
    \caption{Number of unique devices per hour before and after anonymization.}
    \label{fig:devices_per_hour}
    \Description{Comparison of device counts per hour, before anonymization and after anonymization}
    \centering
\end{figure}

However, for higher privacy guarantees, we introduce noise to the \textit{acctstarttime} attribute using a range of [-3,3] hours. This transformation is illustrated in Figure~\ref{fig:devices_per_hour}.
When comparing unique devices per hour, we note that the trade-off becomes more noticeable.
Nevertheless, intraday increases and decreases in the number of unique devices closely resemble the original dataset. This is particularly relevant, as it enables a reliable study of urban dynamics at the hourly level.
It is also important to note the increase in the total number of devices per hour after the transformations, which was caused by the added noise. As each session usually contains multiple connections from a single device and the start times of these connections are shifted, the session is split into multiple connections that are moved to different hours. This inflates the total number of unique hourly devices, as the device in question may not have used the network during the hours to which the connection has been moved. This essentially creates fictitious records for every hour, which further increases user privacy while maintaining high utility due to the similarity in hourly patterns between the original and transformed data.

Despite these noted fluctuations, we highlight that one of the main advantages of the transformations applied to the original data is the fact that the connection-level granularity is maintained, albeit generalized to hours. This granularity at the connection level ensures that the data remains very close to the raw format, enabling the extraction of metrics within flexible time windows. %, as previously described in Section~\ref{sec:data_description}.
Furthermore, the transformed data can safely be used as input for advanced analyses requiring raw data to capture patterns in high-dimensional data, such as machine learning models, while preserving user privacy. The privacy requirement is enforced in the remaining attributes through pseudonymisation mechanisms. This is not demonstrated in the results, but we emphasize that this analysis includes such transformations; as we observe, all patterns are preserved.

\section{Discussion}~\label{sec:discussion}

Using Wi-Fi access data has several advantages. Firstly, Wi-Fi APs are well established in most public buildings in a smart city, meaning there is no need to install additional equipment to collect data. %It's almost free and easy to get movement data.
Secondly, the number of smartphone users with Wi-Fi functionality is increasing, which means that the size of the data set is massive. Thirdly, in a system like \textit{Eduroam}, data can be collected without the consent of the users if it is anonymized. This type of data collection is referred to as ``opportunistic data''~\cite{bonne2013wifipi}.
%When mobile devices turn on Wi-Fi, they transmit Wi-Fi signals to connect to a Wi-Fi AP. When a device connects to an AP, the distance from the device to the AP can be inferred from the device's signal strength. As the device moves, it moves closer to other APs. 
However, by analyzing the Wi-Fi connection data together with the correct location of the APs, the movement of the device can be tracked, and the behavior of people can be further investigated. 
%As such, it is crucial to protect this data. 

Since this type of data is crucial for the continuous development of a smart city, we must avoid this scenario. Therefore, we applied several Privacy-Preserving Techniques (PPTs) to protect all attributes in the dataset. We focused on reducing the granularity of the data, suppressing critical records and pseudonymizing IDs such as MAC addresses. These truthfulness transformations allow to maintain semantic consistency; thus, we increased the indistinguishability of individuals while maintaining interpretability.

Although the MAC addresses of devices (users and APs) are not shared in raw format, there is a possibility that an attacker may be able to infer the location of specific APs from the volume of data by cross-referencing with public information. As a result, the attacker may be able to map some trajectories. For this reason, and to avoid time-related inference information, we have applied noise to the \textit{acctstarttime} attribute by shifting [-3, 3] hours. We aimed to minimize the distortion in order to preserve more utility while enhancing privacy.
Thus, we have increased the uncertainty of the attacker's guess, for which the risk of re-identifying a user is close to zero. Nevertheless, Figure~\ref{fig:devices_per_hour} shows that even with two transformations on the same attribute, utility can be preserved, with the applied PPTs still allowing for posterior data analysis and aggregated information for the metrics.

In general, our results point towards a balance between privacy and data utility. Note that, in our analysis, we do not perform any privacy attack as demonstrated in the relevant literature (e.g.~\cite{ackermann2023privacy,demir2013wi}). Our goal is to develop a methodology to protect the full dataset by performing a general privacy analysis that can be applied by a wide range of individuals, regardless of their background in data privacy or data science and their technical expertise. Therefore, we focused on simple analysis by targeting small groups that could be easily identified.

Despite the successful outcome, there are additional concerns that need to be addressed. Establishing metrics to provide an overall view of network performance may require additional effort, as these metrics are fundamental to capturing usage patterns and detecting anomalies, as well as supporting network management decisions. In this regard, these metrics should be updated regularly for higher control. Thus, the data custodians can build a system in their infrastructure to account for this goal. In particular, create several linked tables that contain the aggregated information, which can be consulted at any time. For each temporal granularity, a new table can be created. Nevertheless, these tables must incorporate privacy mitigations such as removing the devices/sessions lower than a defined threshold. Also, before the pseudonymization of MAC addresses, further protection can be applied, such as removing the last bits. Since these tables may involve a longer period of data, this transformation is crucial to avoid time-correlated inferences. It is essential to perform an analysis in order to evaluate the impact of losing this granularity. For instance, in our context, Table~\ref{tab:cutbytes} shows such an impact considering that the number of initial unique MAC addresses was 5.843.503.

\begin{table}[ht!]
  \caption{Analysis of the impact of removing the last bits from MAC addresses.}
  \label{tab:cutbytes}
  \begin{tabular}{ccc}
    \toprule
    \textbf{Removed bits}&\textbf{Nr. of unique MAC addresses}&\textbf{Impact (\%)}\\
    \midrule
    \textit{1 bit} & 5.806.060 & 0.641\\
    \textit{2 bits} & 5.737.105 & 1.821\\
    \textit{3 bits} & 5.632.920 & 3.604\\
    \textit{4 bits} & 5.496.409 & 5.940\\
  \bottomrule
\end{tabular}
\end{table}

The results clearly indicate a correlation between the removal of more bits and a greater impact on the utility of the data, i.e., fewer distinct observations and, consequently, patterns. The data controller must evaluate the benefits of removing a certain number of bits and choose the transformation with the best trade-off.

A limitation of our methodology is that we focus only on identity disclosure. However, it should be considered whether there are cases of possible attribute disclosure.
Although the main concern of the regulator is to ensure that there is no entity disclosure because of the severe consequences for both individuals and organizations, other metrics should be used to evaluate attribute disclosure, such as $l$-diversity~\cite{machanavajjhala2007diversity} and $t$-closeness~\cite{li2006t}. Although these metrics may help in finding some homogeneity, we stress that in Wi-Fi network data or mobility data in general, they may not be as effective as in static data. Record linkage metric can also be used for %re-identification and
attribute disclosure~\cite{carvalho2021fundamental}, where in this case we suggest using subsets since record linkage requires high computation. %Note that record linkage combines two tables, thereby requiring an external dataset.
%Although the two risk metrics used are applied for different purposes, namely $k$-anonymity on truthful and record linkage untruthful transformations, both metrics should be used simultaneously, since the former gives us the proportion of single outs, whereas the latter provides the proportion of information that is unprotected.

Besides the high importance of Wi-Fi network data, a smart city involves other types of data. In this regard,~\citet{sampaio2023collecting} provides a survey that guides on the processing of personal data along with techniques and tools to (pseudo) anonymize data in smart cities.

%These data support data-driven decision making across various urban systems thus contributing to economic efficiency, social inclusion, and environmental sustainability in smart urban development.

\subsection{Legal Considerations}
Nowadays, both public and private entities have the power to access a significant amount of information about the citizens. For this reason, the right to personal data protection is mainly about giving citizens the tools to control the processing of their data and, consequently, to decide what happens to their information, i.e., informational self-determination. 
In other words, the right to data protection gives individuals the right to know \textit{i)} who is processing their data; \textit{ii)} what data is being processed; \textit{iii)} what the actual purpose of the processing is; \textit{iv)} and to whom it is being transmitted. These culminate in the right to information (Art. 13 and 14 GDPR); the right of access, which someone can actively exercise over any entity (Art. 15 GDPR); and the right to rectification (Art. 16 GDPR). 
Therefore, the right to data protection is not a value in itself, but an instrumental right to guarantee human dignity -- a guarantee of many other fundamental dimensions, such as privacy, freedom, the free development of personality and equality. 
% Moreover, the GDPR aims to implement requirements to protect citizens, in a much like German logic of informational self-determination.

In this section, we discuss and assess the GDPR compliance of the use of Wi-Fi network data, subject to (pseudo) anonymization processing, for sharing such data. The goal is to guide data controllers in complying with their data protection obligations. Typically, this is a process containing the following steps: 
\textit{i)} data minimization and pseudonymization, \textit{ii)} lawfulness of processing, \textit{iii)} data protection by design and default, and \textit{iv)} periodic privacy impact assessments.

%\textit{i)} general description, \textit{ii)} data processing principles, \textit{iii)} lawfulness of processing, \textit{iv)} data protection by design and by default, and \textit{v)} security of processing.

%\paragraph{\textbf{General description}}
%Focusing on a detailed analysis of the process, the first step was to remove from the dataset those attributes that were not considered useful for the intended purpose; as the logs contain multiple attributes, 21 of the 28 available attributes were removed. %, thus respecting the data minimisation principle 
%In addition, the previous process allows for the protection of the full dataset.The data can be considered anonymized insofar as no additional information is shared that would allow the pseudonymization to be reversed. In short, the application of robust pseudonymization measures can substantially reduce the risk of privacy disclosure, but does not completely nullify the possibility, and should therefore be treated as personal data under the GDPR. 

\paragraph{\textbf{Data Minimization and Pseudonymization.}} 
One crucial aspect of GDPR compliance is adhering to the principle of data minimization, as outlined in Article 5(c). This means collecting and processing only the minimum amount of data necessary to achieve the intended purpose. When working with Wi-Fi network data, it is essential to remove any unnecessary attributes that do not serve a specific function. For example, if the aim is to analyse network usage patterns, it may not be necessary to retain all the attributes from the server's log files. In our case, 21 of the 28 available attributes were removed, as they were not relevant to our defined purpose. 
%to retain identifiable user information such as MAC addresses or device serial numbers. 
Reducing the dataset to strictly what is essential not only complies with GDPR requirements but also minimizes the potential risks associated with data breaches.

Pseudonymization further enhances privacy by transforming personal data in a way that makes it difficult to trace back to an individual without additional information. For example, replacing MAC addresses with hashed values can prevent direct identification of users while still allowing for meaningful analysis of network activity. However, it is important to note that pseudonymized data is still classified as personal data under the GDPR unless anonymization is achieved through irreversible methods, such as aggregating data to a point where individuals cannot be singled out.

\paragraph{\textbf{Lawfulness of Processing.}} 
Another critical consideration is establishing a lawful basis for processing Wi-Fi network data. Under Article 6 of the GDPR, one must identify a valid legal ground before initiating any data processing activities. Common bases include obtaining explicit consent from data subjects, pursuing legitimate interests (e.g., improving network performance or security), or fulfilling a contractual obligation. If relying on legitimate interests, it is essential to conduct a balancing test to ensure that the processor's interests do not outweigh the fundamental rights and freedoms of the individuals involved.
Transparency is key in this process. Data subjects should be informed about the collection and use of their data through clear privacy notices. These notices should detail the purposes for which the data will be used, the lawful basis for processing, and any third parties with whom the data may be shared. By ensuring transparency and obtaining proper legal consent where necessary, one can build trust with data subjects while maintaining compliance with GDPR standards.

\paragraph{\textbf{Data Protection by Design and Default.}}
Implementing measures that integrate data protection into the design and by default is a fundamental requirement under Article 25 of the GDPR. This involves embedding privacy considerations at every stage of the processing activities, from initial data collection to final disposal or anonymization. In the context of Wi-Fi network data, this could entail establishing access controls to restrict who can view or manipulate the data, encrypting sensitive information during transmission and storage, and periodically reviewing retention policies to ensure that data is not retained beyond what is necessary.
Alongside technical measures, organizational practices are vital for safeguarding data. Training staff on GDPR principles and establishing clear accountability structures can help prevent unintentional breaches. Adopting a proactive approach to privacy not only meets regulatory expectations but also fosters a culture of data protection within any organization.

\paragraph{\textbf{Periodic Privacy Impact Assessments.}}
To maintain ongoing compliance with the GDPR, it is advisable to conduct periodic Privacy Impact Assessments (PIAs). PIAs are systematic evaluations that help identify and mitigate potential privacy risks associated with processing personal data. Concerning Wi-Fi network data, a PIA might involve assessing how changes in technology or usage patterns could affect privacy protections. It could also explore whether existing pseudonymization techniques remain robust against emerging threats, such as advancements in data re-identification methods.
By regularly reviewing these processes and updating them as needed, one can adapt to evolving regulatory requirements and technological landscapes. This proactive approach not only ensures continued compliance but also demonstrates a commitment to protecting users' privacy, which is increasingly valued in today's digital environment.

%\subsection{Challenges in Network Data Management}
% Distributed MobilityDB: A Scalable Moving Object Database Management System

% Data Architecture and Big Data Analytics in Smart Cities

\subsection{Final Remarks}
Currently, privacy faces three main challenges: unawareness of the importance of protecting private information, lack
of knowledge about privacy preservation methods, and the idea that preserving privacy destroys utility, which has long been
nurtured but needs to be deconstructed. It is crucial to convey that although the process of de-identification requires extra effort,
it is possible to preserve privacy without compromising utility too much. 

Open data has long been used to benefit society in a variety of domains. The pandemic has highlighted the urgency of making data available to the public, and in the context of smart cities, making data available will foster innovation. 
However, several key concerns must be carefully addressed throughout the entire process, from data acquisition to public data release. These concerns typically include: \textit{i)} transparency and consent about the data use, \textit{ii)} data minimization by sharing only what is essential for public benefit, \textit{iii)} privacy prioritization through de-identification techniques, \textit{iv)} ensure that the data remains accurate and unbiased after transformations to prevent misleading insights, and \textit{v)} accountability by documenting the applied methods to align with public trust and data protection laws.

While prioritizing individual privacy is essential, achieving it is a complex and challenging task. Data are often minimized and generalized to the point of being statistically useless. This is a consequence of the misuse of optimal PPTs or high PPT parameters, the use of misleading case studies, disregard of the purpose of releasing the data, and many other causes. This can be circumvented with the participation of a multidisciplinary team.

To conclude, we emphasize that de-identification experts should educate and promote awareness on how to protect private information. Collaboration with data custodians -- individuals well-versed in the data domain -- is essential. During this process, the experts should demonstrate the different stages of the de-identification process, as this is the first step in raising awareness of the workflow involved. 
The synergy between the data protection team, the data release team and the end users is crucial for producing a statistically significant data set with a high level of privacy protection.
Nevertheless, collaborating across different groups may present two main challenges. On the one hand, data custodians may focus on the usability of the data, while the data protection team may prioritize strict privacy measures. On the other hand, demonstrating and validating each step of the de-identification process can be resource-intensive, especially when coordinating multiple groups. Aligning the conflicting priorities and balancing resources and time are critical to successful de-identification.

\section{Conclusion}~\label{sec:conclusion}
Network data is widely used in smart city projects. Wi-Fi networks provide information on human mobility patterns, space utilization, and service demand, providing a rich illustration of how cities, public places, and urban spaces are used. However, raw Wi-Fi network data contains sensitive information, including MAC addresses, timestamps, and location data, which can be used to track individuals. In this paper, we highlight the importance of protecting all attributes by providing a detailed de-identification process. We focus on simple transformations that are effective in preserving privacy without destroying the utility. This process was carried out in collaboration with a multidisciplinary team, which, while often challenging to assemble, is essential to achieving optimal results. Most importantly, we provide legal considerations to translate the current regulatory framework into actionable technical guidelines for responsible data sharing. The resulting (pseudo) anonymized dataset was successfully used in a hackathon, demonstrating the practicality of our methodology in the real world. Finally, we stress the importance of raising privacy awareness by demonstrating effective de-identification procedures to support responsible open data sharing to foster innovation. 

% \begin{acks}
% \end{acks}

\bibliographystyle{ACM-Reference-Format}
\bibliography{sample-acmsmall}

% \appendix



\end{document}